\documentclass[11pt]{article}
\usepackage{english}
\usepackage{A4}
\usepackage{isolatin1}
\usepackage{epsf}
\usepackage[dvips]{graphicx}

\newcommand{\eps}[4]
{
 \begin{figure}[hbt]
  \centering
  \includegraphics[scale=#2,clip=true]{#1}
  \caption{#3}
  \label{#4}
 \end{figure}
}

\begin{document}

\title{LuaJava -- A Scripting Tool for Java}
\author{Carlos Cassino \and Roberto Ierusalimschy \and Noemi Rodriguez \and
{\small e-mail: {\tt cassino,roberto,noemi@inf.puc-rio.br}}}
\date{PUC-RioInf.MCC02/99 February, 1999}

\maketitle

\thispagestyle{empty}

\noindent
Abstract:
Scripting languages are becoming more and more important as a tool for 
 software development, as they provide great flexibility for rapid 
 prototyping and for configuring componentware applications.
In this paper we present LuaJava, a scripting tool for Java.
LuaJava adopts Lua, an dynamically typed interpreted language,
 as its script language. 
Great emphasis is given to the transparency of the integration between the 
 two languages, so that objects from one language can be used inside the 
 other like native objects.
The final result of this integration is a tool 
 that allows the construction of configurable Java applications,
 using off-the-shelf components, in a high abstraction level.

\vspace{0.2in}

\vspace{0.5in}
\noindent
Resumo:
Linguagens de script têm ganho cada vez mais importância no desenvolvimento de
 aplicações pois provêem uma grande flexibilidade para tarefas como programação
 exploratória, prototipação, configuração e construção de aplicações baseada em
 componentes.
Neste artigo apresentamos LuaJava, uma ferramenta de scripting para Java.
LuaJava utiliza Lua---uma linguagem interpretada e dinamicamente tipada---como
 linguagem de script.
Uma grande ênfase é dada à transparência da integração de um objeto Java em Lua
 e vice-versa, de modo a que um objeto externo possa ser utilizado da mesma
 forma que um objeto nativo.
O resultado dessa integração é a possibilidade de se construir aplicações Java
 altamente configuráveis a partir de componentes pré-concebidos, em um elevado
 nível de abstração.

\vspace{0.2in}
\noindent
Palavras-chave: Linguagens de script, Componentes de Software

\vspace{0.5in}
\noindent
This work has been sponsored by CAPES and CNPq.

\pagebreak

\setcounter{page}{1}

\section{Introduction}

Scripting languages, such as Tcl, Lua, VisualBasic, and Python,
have a growing importance
in software development \cite{scripting,vba_and_com}.
Such languages, due to their interpreted and dynamic nature,
stand up well to the tasks of exploratory programming,
prototyping, and configuration of component-based applications.
Development languages such as Java, C, and C++, while suitable
for large system construction, are not ideal for these specific tasks.

In the specific case of a statically typed OO language,
in order to experiment with a solution we must
produce all declared methods and attributes,
even if only a few of them are used in the test.
When using a dynamically typed language, we can work with partial
implementations:
Unused methods and attributes have no influence on test results.
The use of an interpreted language also allows code to be written
directly and interactively, 
abolishing the need for a programming environment and even for a compiler.

The use of multiple programming languages has been widely explored
in software development.
One frequent scenario is the use of two languages,
a \emph{host} language and a \emph{scripting} language.
The host language, usually a compiled language,
is used to write software components,
which are then glued together by the scripting language.
This architecture explores the best features of each kind of language.
Compiled, strongly-typed languages have the robustness
and efficiency that is deemed necessary for component construction.
Interpreted, dynamically typed languages can deal with these components with
no need for previous declarations or for a compilation phase.
Besides, the higher abstraction level typical of scripting languages
allows programs of a few lines to handle tasks which would
demand dozens of lines of code in a language such as C.

In this paper we present LuaJava, a scripting tool for Java.
The goal of this tool is to allow scripts written in the
extension language Lua~\cite{lua,lua-ddj,lua-www} 
to manipulate components developed in Java.
Since the role of the scripting language is to provide
a high level of abstraction for this manipulation,
an essential feature in the tool is its ease of use.
LuaJava allows Java components to be accessed from Lua scripts
using the same syntax that is used for accessing Lua's native objects,
without any need for declarations or any kind of preprocessing.

A good scripting tool must allow us not only to
manipulate components previously written in the host language,
but also to define new components compatible with the host language.
The use of the callback mechanism is a common example of this need;
in a purely OO language,
a callback is represented by an object whose 
methods are called to handle the monitored events.
To code a callback button in a graphic interface,
we must create a new class, compatible with 
the type expected for button callbacks,
containing the desired code. 
Callback mechanisms are common for a myriad of
tasks besides graphical interface control.
LuaJava fulfills this requirement by allowing
integration in both directions:
manipulation of Java objects by Lua scripts and manipulation of
Lua objects by Java programs.

Lua, like most scripting languages, is not untyped,
but strongly dynamically typed.
That means that you cannot execute an operation over an invalid type,
but such errors are detected only at run time.
In order for the integration of Lua with Java
to be coherent with the dynamic nature of the scripting language,
the scripting tool must be able to do dynamic type checking.
Each time the scripting language access a host component,
the tool must check the existence of the required operation,
the correctness of the arguments, etc.
The scripting tool makes extensive use of Java's
reflexive API to carry out these checkings.
This binding model, based on reflexivity and dynamic typing,
can be contrasted with more traditional bindings between a scripting
language and a host language,
where a new stub must be created and compiled
for each new component to be integrated.

In the next section we present the LuaJava tool;
next we discuss its implementation in Section~\ref{impl}.
We compare LuaJava with some related work at Section~\ref{trab},
where we also show some performance measures. 
As usual, we devote the last section for some conclusions and
final remarks.

\section{The LuaJava Tool}

One of the goals of LuaJava is to allow the programmer to manipulate
Java objects in the same way as she manipulates native (Lua) objects.
So, first let us see how Lua objects are manipulated.

Lua, like most interpreted languages, is dynamically typed.
Variables have no type; instead, each value carries its own type with it.
Lua has no declarations;
any variable may contain any value of the language.
For instance, to create a new object modeling a two
dimensional point in Lua,
we may just write
\begin{verbatim}
point = {x=0, y=0}
\end{verbatim}
That will create a new object,
with two fields \verb|x| and \verb|y|,
both with initial values of 0.
Later, we can add other fields to this object,
for instance a field containing a function (a method):
\begin{verbatim}
function point:move (dx,dy)
  self.x = self.x + dx
  self.y = self.y + dy
end
\end{verbatim}
This code creates a new field, called \verb|move|,
whose value is a function with three parameters:
a hidden \verb|self|, \verb|dx| and \verb|dy|.
This method can then be called with the following syntax:
\begin{verbatim}
      point:move(2,3)
\end{verbatim}
which is simply syntactic sugar for \verb|point["move"](point,2,3)|.
In other words, the field indexed by the string \verb|"move"| is
accessed, and its value is called with the object itself as
its first hidden argument
(if the value of \verb|point["move"]| is not a function we get a
run time error).

The lack of type declarations and the possibility of
dynamic creation of methods facilitate the integration
of Java objects into Lua scripts.
Based on Lua's dynamic nature, LuaJava allows an external object,
whose type is unknown up to the moment it is accessed,
to be manipulated as if it were a native Lua object.
As an example, consider the following Java class:
\begin{verbatim}
public class Point {
  public float x=0, y=0;
  public void move(float dx, float dy) {
    x+=dx; y+=dy;
  }
}
\end{verbatim}
In order to instantiate this class in a Lua script,
we can use the {\tt javaNewInstance} function,
defined by LuaJava:
\begin{verbatim}
point = javaNewInstance("Point") -- point is now a proxy to a Java object.
\end{verbatim}
This function creates a new Java object,
and returns a Lua object that is a proxy to the Java object.
Now we can access this object with the same syntax previously
used for the Lua implementation of a point.
The following script moves the point in two different ways, 
first by invoking method {\tt move}, and then by directly
accessing its attributes.
The script then prints the new values of the point's coordinates
using Lua's {\tt print} function.
\begin{verbatim}
point:move(2,3)

point.x = point.x+1
point.y = point.y+1

print(point.x, point.y)
\end{verbatim}

As this example shows, manipulation of Java objects with LuaJava
is {\em transparent} to the Lua programmer:
the syntax used for manipulating Java objects is exactly the same as
that used for manipulating a native Lua object.

Transparency is important for a number of reasons.
The first of them is ease of use; 
it is important that the programmer should not have
to learn a third syntax, besides those of Lua and of Java:
If a Java object is being handled in Lua, then the Lua syntax is used.
Another reason is to allow existing Lua scripts to work upon
Java objects;
if a specific syntax was used, this integration would not be direct.
Finally, transparency allows Lua to be used as a framework for
component integration.
Given this feature, it is possible to write scripts that
handle objects regardless of their origin (C++, Java, etc.).

Besides {\tt javaNewInstance},
LuaJava defines function {\tt javaBindClass},
which retrieves a Java class.
The returned object can be used to access fields and static methods
of the corresponding class.

As another, more interesting, example of the use of LuaJava,
we will outline the construction of
a console where the user can interactively write and execute
code in the scripting language.
Such a console allows the user direct run time control over the
application objects.
To build it, we will use a Lua script and some interface elements from
the standard {\tt java.awt} package~\cite{awt}.
\begin{figure}
\small
\begin{verbatim}
window = javaNewInstance("java.awt.Frame", "Console Lua AWT")
text = javaNewInstance("java.awt.TextArea")
button = javaNewInstance("java.awt.Button", "Execute")

BorderLayout = javaBindClass("java.awt.BorderLayout")

window:add(text, BorderLayout.NORTH) 
window:add(button, BorderLayout.SOUTH)
window:pack()
window:show()
\end{verbatim}
\caption{Creating a Console.\label{console}}
\end{figure}
Figure~\ref{console} presents all the code we need to show the console.
This console, however, lacks functionality.
When its button is pressed, nothing happens.

To add functionality to our console,
we must create a \emph{listener} for its button,
that is, a callback object with a method to be called when the
button is pressed.
Instead of creating a Java object to be manipulated by Lua,
now we want to create a Lua object that will be manipulated by Java.
We can do that in LuaJava by simply creating the Lua object and using
it as an argument where a Java object is expected.
In our example, the following code is enough:
\begin{verbatim}
button_cb = {}

function button_cb:actionPerformed(ev)
  dostring(text:getText())
end

button:addActionListener(button_cb)
\end{verbatim}
In the first line we create an empty Lua object;
next, we define a new method for this object, \verb|actionPerformed|.
This method, when invoked, executes the text contained
in the \verb|TextArea| element as a piece of Lua code.
Finally, in the last line, LuaJava does the main work for us.
When we pass this object to the method \verb|addActionListener|,
which expects a listener,
LuaJava detects this and automatically produces
a Java wrapper for this object.
This wrapper is then used as the actual listener to be installed.
Whenever method \verb|actionPerformed| is called,
the wrapper calls the corresponding method in the Lua object.

Although in our example the Lua object is used as a listener,
LuaJava is not restricted to these cases. 
We can use Lua scripts to write implementations for any type in Java,
both interfaces and classes.
Usually this is done automatically,
if a Lua object is used where a Java
object is expected.
The programmer may also explicitly create a wrapper for a Lua object,
using function \verb|javaExport|.

\section{The Implementation of LuaJava} \label{impl}

LuaJava is implemented through \emph{proxies}:
whenever a Java object is passed to Lua,
a Lua object is created to represent it.
Conversely, when a Lua object is passed to Java,
a Java object is used to represent it.
Because Lua and Java have quite different run time behaviors,
we have used different techniques to implement the Lua proxies and
the Java proxies.

A Lua proxy (that is, a Lua object that represents a Java object)
is implemented using the \emph{fallback} mechanism~\cite{lua}.
A fallback, in Lua, is a function that can be used to modify the
default behavior of the language for some events.
In LuaJava we use fallbacks to modify the way proxy
fields are accessed and modified.
So, a proxy is a nearly empty Lua object that
intercepts any access to its (usually empty) fields.
A proxy has also at least one non empty field,
that stores a reference to the Java object it represents.

Figure~\ref{method-call} shows what happens when
we call a proxy's method:
\eps{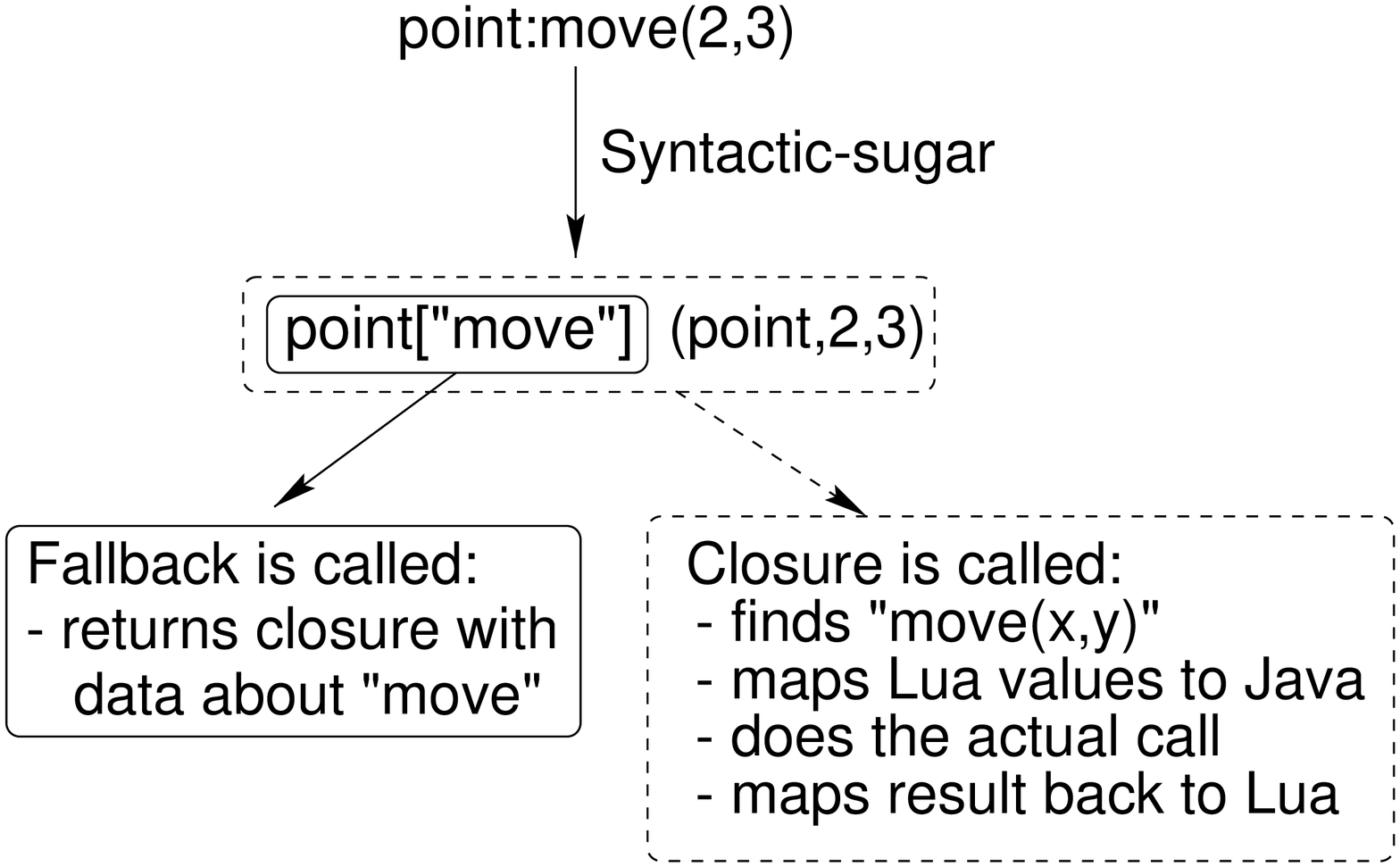}
    {0.4}
    {Calling a Java method from Lua.}
    {method-call}
The first step, shown in the solid box,
is the access to a field called \verb|move|.
This access triggers a fallback;
the fallback, having access to the proxy
(and thus to the Java object it represents) and the method name,
uses Java's reflexive API to obtain information about that method
(or methods, since Java allows overloading).
The fallback then returns a \emph{closure} with this information.%
\footnote{A closure is a a function plus an environment whence
the function gets its non-local variables.}

The second step, shown in the dashed box,
is triggered when the closure is applied to the given arguments.
At this point the closure selects a method suited to the actual parameters,
converts them from Lua to Java,
executes the actual invocation, and converts back the result, if there is one.
As a performance improvement, at the end of the invocation
a closure with the selected method is stored in field 
\verb|move| of the proxy.
Therefore, in the following calls the first step can be avoided.

This same algorithm, with small modifications,
is used to access instance variables and array elements.
Similarly, the fallback over field modifications is used to
handle assignments to array elements and instance variables.

The implementation of Java proxies for Lua objects
is a little more complex.
As Java has type declarations,
each Java proxy must have a predefined type.
Moreover, because Java is statically typed,
we must have a class for each type of proxy that we create.
LuaJava builds these classes on demand.
First, it writes the class bytecode in a byte array,
Next, it uses a \verb|ClassLoader| to dynamically load this new
class into the program.

A Java type can be an interface or a class.
The proxies for these two kinds of types are slightly different.
In both cases, the proxy has an instance variable with a reference to
the Lua object it represents.
For a proxy that implements an interface, LuaJava create a new class,
compatible with the given interface,
with all its methods coded as follows:
First, they convert their arguments to Lua;
next, they call the corresponding method in the Lua object;
Finally, they convert eventual results back to Java.

When the proxy represents a class,
LuaJava implements it as a subclass of the original class.
This subclass redefines all the original methods;
each new method, when called, first checks whether
the corresponding method in Lua exist.
If the Lua method is present,
it is called, like in the case of interfaces.
However, if the Lua object has no such method,
the subclass method detects this absence,
and calls instead its \verb|super| implementation,
that is, the method from the original class.
In this way, a Lua object implementing a class ``inherits''
from the class any method that it does not redefine.

A Lua object representing a Java interface does not need
to implement all its methods, quite the opposite.
After all, one of the goals of a scripting language
is to allow rapid prototyping.
With LuaJava, we need to implement
only the methods that will be used.
Moreover, we can add methods to our object incrementally,
even after it has been passed to Java.

\section{Related Work} \label{trab}

Several other bindings between scripting languages and Java
have been proposed.

For Tcl~\cite{tcl} two integration solutions exist:
the TclBlend binding~\cite{tclblend} and the Jacl implementation~\cite{jacl}.
TclBlend is a binding between Java and Tcl, which, as LuaJava,
allows Java objects to be manipulated by scripts.
Some operations,
such as access to fields and static method invocations,
require specific functions.
Calls to instance methods are handled naturally
by Tcl commands.

TclBlend also allows the creation of callbacks in Tcl,
but only for listeners that follow the JavaBeans protocol.
This leads to a certain lack of generality,
as the programmer can create implementations only of Java types
that are defined as listeners.
Call-back handlers in TclBlend scripts are thus restricted to
components compliant with the JavaBeans standard.

The other integration solution for Tcl and Java,
Jacl, is a new Tcl interpreter written in Java,
which incorporates the TclBlend facilities.
Thus, Jacl acts as a 100\% pure Java substitute for TclBlend,
and it works even when support for native methods is not available.

The language Python~\cite{python} also
has been integrated with Java through a
new interpreter entirely written in Java,
called JPython~\cite{jpython}.
Like the other tools, JPython allows Java objects to be manipulated like
native (Python) objects in a script.
Moreover, Python objects can also be used to implement Java objects,
provided that they have compatible types.

Perl also can be used as a scripting language for Java.
The tool JPL~\cite{jpl} has a preprocessor that extends the Java syntax,
so that a class may contain methods written directly in Perl.
However, as the output of the preprocessor
must be compiled before being used,
most of the advantages of a scripting language are lost.
Another tool for Perl, Jperl~\cite{jperl}, provides a low level binding,
which allows Java to run Perl scripts.

The approach of reimplementing the scripting language in Java,
adopted by Jacl and JPython,
has some drawbacks, besides the work it involves.
First, it is very difficult to keep both
implementations fully compatible.
JACL already has a list of differences from Tcl.
The JPython documentation, on the other hand, says that such a list will
be available as soon as the new implementation becomes stable.

Second, a Java implementation aggravates the performance penalty of
using a scripting language.
\begin{table}
  \centering
  \begin{tabular}{|l||r|r|} \hline \hline
             & Original & Java Imp.         \\ \hline \hline
    Python   & 15s      & $12\times10^1$s \\ \hline
    Tcl      & 46s      & $15\times10^2$s \\ \hline \hline
  \end{tabular}
  \caption{Time to run a simple test.\label{tab1}}
\end{table}
Table~\ref{tab1} shows the time required to run
a script that calls $10^6$ times
a method that increments a variable.

The performance of LuaJava, although not very good,
is compatible with the performance of similar tools.
Table~\ref{tab3} shows the time required to make inter-language calls
in those tools.
\begin{table}
  \centering
  \begin{tabular}{|l||r|r|} \hline \hline
    SL      & SL$\rightarrow$Java & Java$\rightarrow$SL \\ \hline \hline
    LuaJava & $49\mu$s            & $64\mu$s            \\ \hline
    JPython & $54\mu$s            & $67\mu$s            \\ \hline
    JACL    & $475\mu$s           & $672\mu$s           \\ \hline \hline
  \end{tabular}
  \caption{Performance of inter-language calls.}
  \label{tab3}
\end{table}
This table was computed by running three loops of $10^6$ iterations,
the first one empty, the second calling an empty method,
and the third calling the empty method twice.
The time shown in the table is the average of the differences between
those loops divided by the number of iterations.
All measures were made in a Pentium 200Mhz, under Linux.

The time required for a method call inside Lua is approximately $3\mu$s,
while the time to call a Java method from Lua is $49\mu$s.
This tenfold increase is due to the dynamic behavior of such calls,
plus the conversion of arguments (including \emph{self})
and results from one language to the other.

\section{Final Remarks}\label{conc}

We have achieved a good degree of integration between Java and
the scripting language Lua.
Java objects can be easily manipulated by Lua scripts,
and Lua objects can be used to implement both Java interfaces and classes.
Following the goals of a scripting language,
the binding provided by LuaJava is fully dynamic.
Type checking and bindings are all done at invocation time.
Therefore, you can build and modify your scripts incrementally,
without the need to rebuild the system after each modification.
Also, we can provide objects with partial implementations,
without stubs for absent methods;
each object may get new methods and instance variables on demand,
even after its bind to Java.

LuaJava is part of a larger project, called LuaOrb~\cite{luaorb,ry98-1}.
LuaOrb integrates a Lua script not only with Java components,
but also with CORBA and COM components.
Due to the systematic use of the reflexive techniques presented here,
all those bindings offer the same facilities that LuaJava.
Moreover, the objects created by the bindings are fully compatible.
For instance, it is possible to use a CORBA object as a
listener for a Java component.

\nocite{java}
\nocite{java-spec}

\newcommand{\etalchar}[1]{$^{#1}$}


\begin{thebibliography}{WPS{\etalchar{+}}97}

\bibitem[AG97]{java}
Ken Arnold and James Gosling.
\newblock {\em The {Java} Programming Language, 2nd Edition}.
\newblock Addison-Wesley, 1997.

\bibitem[Bal99]{jperl}
S.~Balamurugan.
\newblock {Jperl}: Accessing {Perl} from {Java}.
\newblock {\em Dr. Dobb's Journal}, February 1999.

\bibitem[CIR98]{luaorb}
Renato Cerqueira, Roberto Ierusalimschy, and Noemi Rodriguez.
\newblock Dynamic configuration with {CORBA} components.
\newblock In {\em Fourth International Conference on Configurable Distributed
  Systems}, pages 27--34, May 1998.

\bibitem[CL98]{awt}
Patrick Chan and Rosanna Lee.
\newblock {\em The {Java} Class Libraries, 2nd Edition, Volume 2}.
\newblock Addison-Wesley, 1998.

\bibitem[FIC96]{lua-ddj}
L.~H. Figueiredo, R.~Ierusalimschy, and Waldemar Celes.
\newblock {Lua}---an extensible embedded language.
\newblock {\em Dr. Dobb's Journal}, 21(12):26--33, 1996.

\bibitem[Gat98]{vba_and_com}
Bill Gates.
\newblock {VBA} and {COM}.
\newblock {\em Byte Magazine}, 23(3):70--72, March 1998.

\bibitem[GJS96]{java-spec}
James Gosling, Bill Joy, and Guy Steele.
\newblock {\em The {Java} Language Specification}.
\newblock Addison-Wesley, 1996.

\bibitem[ICR98]{ry98-1}
Roberto Ierusalimschy, Renato Cerqueira, and Noemi Rodriguez.
\newblock Using reflexivity to interface with {CORBA}.
\newblock In {\em {IEEE} International Conference on Computer Languages
  ({ICCL}'98)}, pages 39--46, Chicago, {IL}, May 1998. IEEE Computer Society.

\bibitem[IdFC96]{lua}
R.~Ierusalimschy, L.~H. de~Figueiredo, and W.~Celes.
\newblock {Lua}---an extensible extension language.
\newblock {\em Software: Practice \& Experience}, 26(6):635--652, 1996.

\bibitem[IdFC98]{lua-www}
R.~Ierusalimschy, L.~H. de~Figueiredo, and W.~Celes.
\newblock {\em Reference Manual of the Programming Language {Lua} 3.1}.
\newblock {TeCGraf}, Computer Science Department, {PUC-Rio}, 1998.
\newblock {\tt http://www.tecgraf.puc-rio.br/lua/manual}.

\bibitem[Joh98]{jacl}
Ray Johnson.
\newblock {Tcl} and {Java} integration, January 1998.
\newblock {\tt http://sunscript.sun.com/java/tcljava.ps}.

\bibitem[Lut96]{python}
Mark Lutz.
\newblock {\em Programming {Python}}.
\newblock O'Reilly \& Associates, 1996.

\bibitem[Ous94]{tcl}
John Ousterhout.
\newblock {\em {Tcl} and {TK} Toolkit}.
\newblock Addison-Wesley, 1994.

\bibitem[Ous98]{scripting}
John Ousterhout.
\newblock Scripting: Higher level programming for the 21st century.
\newblock {\em IEEE Computer}, 31(3):23--30, March 1998.

\bibitem[Sta98]{tclblend}
Scott Stanton.
\newblock {TclBlend}: Blending {Tcl} and {Java}.
\newblock {\em Dr. Dobb's Journal}, pages 50--54, February 1998.

\bibitem[vR98]{jpython}
Guido van Rossum.
\newblock {Java} and {Python}: a perfect couple.
\newblock {\em developer.com}, August 1998.
\newblock {\tt http://www.developer.com/journal/techfocus/}
  {\tt 081798\_jpython.html}.

\bibitem[WPS{\etalchar{+}}97]{jpl}
Larry Wall, Nate Patwardhan, Ellen Siever, David~Futato \, and Brian Jepson.
\newblock {\em {Perl} Resource Kit -- UNIX Edition}.
\newblock O'Reilly \& Associates, 1997.
\newblock {\tt http://www.oreilly.com/catalog/prkunix/excerpt/UGtoc.html}.

\end{thebibliography}
\end{document}